\documentclass[usenatbib]{mn2e}

\usepackage{graphics}
\usepackage{epsfig}
\usepackage{natbib}

\begin{document}

\title[Starburst cycles in galaxies]
{Quantitative constraints on starburst cycles in galaxies with stellar masses in the range $10^8-10^{10} M_{\odot}$} 

\author [G.Kauffmann] {Guinevere Kauffmann \\
\thanks{E-mail: gamk@mpa-garching.mpg.de}
Max-Planck Institut f\"{u}r Astrophysik, 85741 Garching, Germany\\}

\maketitle

%===================================
\begin{abstract} 
We have used  4000 \AA\ break and H$\delta_A$
indices in combination with SFR/$M_*$ derived from emission line flux
measurements, to constrain the recent star formation histories
of galaxies with stellar masses in the range $10^8-10^{10} M_{\odot}$.
The fraction of the total SFR density in galaxies with ongoing bursts
is a strong function of stellar mass, declining from 0.85 at a stellar mass of $10^8
M_{\odot}$ to 0.25 for galaxies with $M_* \sim 10^{10} M_{\odot}$. 
Low mass galaxies are not all young. The distribution of half mass
formation times for galaxies with stellar masses less than $10^9 M_{\odot}$ is
broad, spanning the range  1-10 Gyr.
The peak-to-trough variation in star formation rate among the
bursting population ranges lies in the range 10-25. In low mass galaxies, the
average duration of the burst bursts is comparable to the
dynamical time of the galaxy.
Galaxy structure is correlated with estimated burst mass fraction, but in
different ways in low  and  high mass galaxies. High
mass galaxies with large burst mass fractions  are more centrally
concentrated, indicating that bulge formation is  at work. 
In low mass galaxies, stellar surface densities $\mu_*$ decrease as a function of $F_{burst}$.
These results are in  good agreement with the observational
predictions of Teyssier et al (2013) and lend further credence to the idea
that the cuspy halo problem can be solved by energy input from multiple
starbursts over the lifetime of the galaxy.
We note that there is
no compelling evidence for IMF variations in the population of star-forming
galaxies in the local Universe.

\end{abstract}

\section{Introduction}

Cosmological simulations of the evolution of cold dark matter (CDM) show
that the dark matter in collapsed, virialized systems forms cuspy
distributions with inner profiles that are too steep compared with
observations (e.g. Flores \& Primack 1994; Moore 1994; Navarro, Frenk \&
White 1996). This is commonly referred to as the cuspy halo problem.  One
solution to this problem  proposed early on, is that that impulsive
mass loss from the galaxy can lead to irreversible expansion of the orbits
of stars and dark matter near the center of the halo (Navarro, Eke \& Frenk
1996; Read \& Gilmore 2005; Pontzen \& Governato 2012). These conclusions
were based on simple analytic arguments and it was not clear whether this
mechanism could in fact produce central density profiles in close agreement
with observations.  Recently, there have been a series of gas-dynamical
simulations of dwarf galaxies demonstrating that repeated gas outflows
during bursts of star formation can indeed transfer enough energy to the
dark matter component to flatten ``cuspy'' central dark matter profiles
(Mashchenko, Wadsley \& Couchman 2008; Governato et al 2010; Governato et al
2012; Zolotov et al 2012; Trujillo-Gomez et al 2013; Teyssier et al 2013;
Shen et al 2013).

It is still unclear  whether the energy requirements for flattening cuspy
profiles  are in line with the actual stellar populations and star
formation histories of real low mass galaxies (Gnedin \& Zhao 2002;
Garrison-Kimmel et al 2013). In a recent paper, Teyssier et al (2013) highlighted two
key observational predictions of all simulations that find cusp-core
transformations: i) low mass galaxies have a bursty star formation history
with a peak-to-trough ratio of 5 to 10, ii) the bursts occur on the
dynamical timescale of the galaxy.  In addition, the simulations show that
the feedback associated with the bursts alter the final stellar structure
of the galaxies. In simulations without feedback, a prominent bulge with
large Sersic index and a weak exponential disk was formed. In contrast, in
the simulation with feedback, a smaller exponential disk with no bulge in
the center and a clear signature of a core within the central 500 pc was
produced.

Observationally, it is well established that there exist classes of low
mass galaxy, for example blue compact dwarf galaxies with strong emission
lines (BCD galaxies), which are clearly undergoing strong starbursts at the
present day (e.g. Searle \& Sargent 1972; Meurer et al 1995). In very
nearby galaxies where individual stars can be resolved in high quality
images, star formation histories for individual systems can be derived over
longer timescales. Some nearby dwarf spheroidals have clearly experienced
bursty star formation histories in their past (e.g. Dolphin 2012; De Boer
et al 2012). On the other hand, the star formation histories of dwarf
irregular galaxies have been found to be relatively quiescent (e.g. Van Zee
2001).

Recent work on low mass galaxies at high redshifts by Van der Wel et al
(2011)  has indicated that the space densities of extreme emission line
dwarf galaxies with stellar masses $\sim 10^8 M_{\odot}$ are much higher at
earlier cosmic times. At $z>1$, these authors claim that the population of
extreme emission line galaxies detected in the CANDELS survey is high
enough to produce a significant fraction of the total stellar mass
contained in the present-day population of such galaxies, if star formation
continued in the same mode for around 4 Gyr. These authors propose that
most of the stars in present-day dwarf galaxies must have formed in strong
bursts at $z>1$.  This conclusion is, however, at odds with the observation
that the stellar populations of present-day low mass galaxies are, on
average, quite young (Kauffmann et al 2003b). Heavens et al (2004) analyzed
the star formation histories of stacked samples of galaxies in bins of
stellar mass from the Sloan Digital Sky survey. They showed that the {\em
average} star formation history of galaxies with stellar masses less than
$10^{10} M_{\odot}$ could be well represented by flat star formation rate
over the past 3-7 Gyr, with a decline at earlier epochs.

In order to estimate the duty cycle of the starburst phenomenon as well as
the amplitude range in star formation during a burst, it is necessary to
analyze a complete sample of galaxies that are intrinsically similar.
Kauffmann et al (2003b) used constraints from 4000 \AA\ break and
H$\delta_A$ stellar absorption line indices to show that the fraction of
present-day galaxies that have experienced bursts in the past
1-2 Gyr, is a factor of 5 higher for galaxies with stellar masses $\sim
10^8 M_{\odot}$ than for galaxies with $M_* \sim 10^{10} M_{\odot}$. A
similar conclusion was recently reached by Bauer et al (2013), who analyzed
the distribution of specific star formation rates as a function of stellar
mass for galaxies with $0.05 < z < 0.32$  from the Galaxy and Mass Assembly
(GAMA) survey and found a significant tail of galaxies with high SFR/M$_*$
and  low stellar masses that could only be explained by a stochastic burst
with late onset.

In this paper, we use the 4000 \AA\ break and H$\delta_A$ indices in
combination with  SFR/$M_*$ derived from emission line flux measurements,
to constrain the recent star formation histories of low mass galaxies. We
show that this combination of quantities allows us to clearly separate
galaxies that are {\em currently} undergoing a burst of star formation from
galaxies that have formed their stars continuously and galaxies that have
experienced a burst in the past.  We use the subsample of galaxies
experiencing an ongoing burst to constrain burst mass fractions and duty
cycles, as well as the amplitude variation in SFR/M$_*$ during the burst.
We investigate how these parameters change as a function of galaxy mass.
Finally, we investigate how the stellar structure of the galaxy correlates
with burst mass fraction. The paper is organized as follows: in section 2,
we describe or sample and the methodology we use to analyze star formation
histories. In section 3, we present our results, and in section 4, we
summarize and conclude. A Hubble constant $H_0=70$ km s$^{-1}$ is adopted
throughout.

\section {Data Analysis Methodology}

\subsection {The Sample} The parent galaxy sample used in this study is a
magnitude-limited sample constructed from the final data release (DR7;
Abazajian et al. 2009) of the SDSS (York et al. 2000) and is the same as
that used in Li et al. (2012).  This sample contains 482,755 galaxies
located in the main contiguous area of the survey in the northern Galactic
cap, with $r < 17.6$, $24 < M_{0.1r} < 16$ and spectroscopically measured
redshifts in the range $0.001 < z < 0.5$.  Here $r$ is the $r$-band
Petrosian apparent magnitude, corrected for Galactic extinction, and
$M_{0.1r}$ is the $r$-band Petrosian absolute magnitude, corrected for
evolution and K-corrected to its value at z=0.1.

Stellar masses, D$_n$(4000) and H$\delta_A$ line index measurements 
from the SDSS fiber spectra, as
well as the estimated error on these quantities,  are taken from the DR7
MPA/JHU value added catalogue (http://www.mpa-garching.mpg.de/SDSS/DR7/).

Star formation rates estimated within a 3 arcsecond fiber aperture are also available
from these catalogues. The reader is referred to Brinchmann et al. (2004)
for a detailed description of how SFRs are derived. Briefly, star formation
rates are estimated by fitting a grid of photo-ionization models to  the
observed [OIII], H$\beta$, H$\alpha$ and [NII] line strengths for galaxies
where the [OIII]/H$\beta$ and [NII]/H$\alpha$ line ratios have values that
place them within the region of the Baldwin, Phillips \& Terlevich (1981,
BPT) diagram occupied by galaxies in which the primary source of ionizing
photons is from HII regions rather than an active galactic nucleus (AGN).
Standard Bayesian estimation methodology is used to derive the probability
distribution function (PDF) of the star formation rate. We adopt the median
of the integrated PDF as our actual estimate and one half the difference
between the 84th and 16th percentile points as our estimate of the
1$\sigma$ error.  We note that if the H$\beta$ line is not detected, the
[NII]/H$\alpha$ ratio can still be used to separate out AGN; in this case,
the SFR estimates have much larger errors, because dust extinction is not
well constrained.

The MPA/JHU catalogue does contain SFR estimates for AGN and galaxies where
the H$\alpha$ and/or [NII] lines are not detected. These estimates are
derived using an empirical calibration between D$_n$(4000) and SFR/$M_*$
(see Brinchmann et al 2004 for details). For such objects, the SFR
estimates carry no direct information about HII region line luminosities
and thus cannot be used to find galaxies that are experiencing ongoing
bursts. For this reason, we restrict the discussion in this paper to
galaxies with $M_* < 10^{10} M_{\odot}$, where the fraction of AGN drops to
values close to zero (Kauffmann et al 2003c). In this case, essentially all
galaxies with indirect SFR estimates are those with no detection of
H$\alpha$/[NII].  These galaxies have old stellar populations, and as will
be discussed in the next section, they are classified as having had
quiescent star formation histories.

\subsection {Methodology for constraining star formation histories}

We create three different model libraries by using the population synthesis
code of Bruzual \& Charlot (2003) as follows: \begin {enumerate} \item {\em
Continuous Models.} -- a library of model galaxies with continuous star
formation histories.  These model galaxies have a continuous SFR, declining
exponentially according to SFR(t) $\propto e^ {-\gamma t}$, with $\gamma$
uniformly distributed between 0 (i.e. constant star formation rate) and 0.5
Gyr$^{-1}$.  Stars begin to form at a look-back times between 11.375 Gyr
and 4.2 Gyr in the past. We output the spectral energy distribution
parameters at the present day.  \item {\em Models with ongoing bursts.} For each
model galaxy in the continuous library described above, we add a burst
which starts 100 million years before the present time. The burst has
constant amplitude as a function of time and the amplitudes are allowed to
vary so that the mass fraction of stars formed in the burst ranges from
0.00001  to close to 1.  We output the spectral energy distribution
parameters at time intervals $\Delta t$ of 10 million years while the
burst is in progress.  \item {\em
Models with past bursts.} We follow the same algorithm used to create the
models with ongoing bursts, except that the starting time for the burst
ranges from 2 Gyr before the present to 0.2 Gyr before the present. All
bursts run for a total duration of 100 million years. We output
spectral energy distribution parameters at the present day.  \end {enumerate}
We generate all model libraries at metallicities of 0.5 solar and
0.25 solar, appropriate for galaxies with stellar masses in the range
$10^8$ to $10^{10} M_{\odot}$ (Tremonti et al 2004). A fixed Chabrier
(2003) initial mass function (IMF) is adopted for all the models.

The first three panels of Figure 1 show the distribution of model galaxies
in the plane of specific star formation rate (SFR/M$_*$) versus
D$_n$(4000), H$\delta_A$ versus D$_n$(4000), and SFR/M$_*$ versus
H$\delta_A$. Model galaxies in the continuous libraries are coloured in
green, those in the ongoing burst library and coloured in blue and those in
the past burst library are coloured in red. As can be seen, the models with
ongoing and past bursts are well separated, particularly in the SFR/M$_*$
versus D$_n$(4000) and the SFR/M$_*$ versus H$\delta_A$ planes. Continuous
models occupy a narrow locus between the region of the diagrams spanned by
the two classes of burst models.  The locus is particularly narrow for
galaxies with young stellar populations (D$_n$(4000) $ < 1.5$,
H$\delta_A>3$). The continuous models fan out at older ages, where
metallicity affects the 4000 \AA\ break strength (Kauffmann et al 2003a) and
the choice of formation time limits the age of the oldest stars in the
model galaxies.

\begin{figure}
\includegraphics[width=88mm]{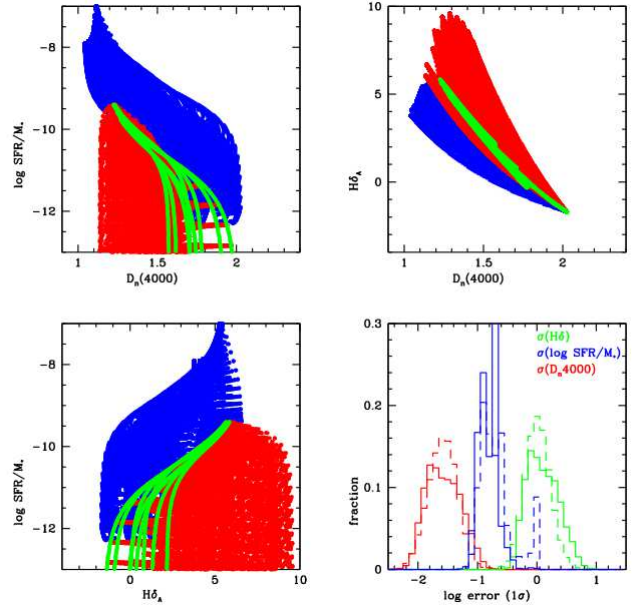}
\caption{ The first three panels  show the distribution of model galaxies
in the plane of specific star formation rate (SFR/M$_*$) versus
D$_n$(4000), H$\delta_A$ versus D$_n$(4000), and SFR/M$_*$ versus
H$\delta_A$. Model galaxies in the continuous libraries are coloured in
green, those in the ongoing burst library and coloured in blue and those in
the past burst library are coloured in red.
In the bottom right panel,
we plot histograms of the 1$\sigma$ errors on D$_n$(4000),
H$\delta_A$ and  SFR/M$_*$.
\label{models}}
\end{figure}

Our ability to use the model library to constrain star formation histories
will depend on the precision to which the quantities D$_n$(4000),
H$\delta_A$ and SFR/$M_*$ can be measured. In the bottom right panel of
Figure 1, we plot histograms of the 1$\sigma$ errors on D$_n$(4000),
H$\delta_A$ and  SFR/M$_*$.  Solid histograms show results for low mass
galaxies ($\log M_* \sim 8.5$), while dashed histograms show results for
high mass galaxies ($\log M_* \sim 10$).  The error distribution depends
very little on mass. The typical error on D$_n$(4000) is a few percent,
while the errors on H$\delta_A$ are around 10\%.  log SFR/M$_*$ is measured
with a formal error of 0.1-0.2 dex. We note that the SFR/M$_*$ error estimate
does not include any estimate of the systematic errors in the photo-ionization
models used to estimate SFR. Nevertheless, it is clear from Figure 1 that
we should be able to separate galaxies with ongoing starbursts from the
rest of the population with reasonably high confidence.

\subsection {Classification and Parameterization of Star Formation History}
We now describe the procedure used to group galaxies into classes according
to whether they  most likely have had continuous star formation histories,
are currently experiencing a burst, or have experienced a burst in the past
2 Gyr. We note that these classifications pertain to the {\em central regions}
of the galaxies, since the fiber spectrum typically samples 20-30 \% of
the total light from the galaxy (Kauffmann et al 2003a).  
We first search the continuous star formation library for the model
that minimizes $\chi^2$. If the minimum $\chi^2$ per degree of freedom
($\chi^2_{min}/N_d$, with $N_d=3$ in our case)  is less than 2.37, this
means that there is a 50\% or greater probability that that the continuous
SFH hypothesis is correct, and we place the galaxy into the continuous
class.

If $\chi^2_{min}/N_d$ is greater than 2.37, we then search both the ongoing
and past burst libraries for a new minimum $\chi^2$ model.  We check
whether the new $\chi^2_{min}/N_d$ is smaller than that obtained for the
continuous library, and smaller than 2.37. If so, we consider our
classification into  ongoing burst/past burst as ``secure''.  We note that
the burst libraries contain models with burst fractions very close to zero,
which are essentially identical to the models in the continuous library, so
the new minimum $\chi^2$ is guaranteed to be equal to or smaller than the
old one.  If the new $\chi^2_{min}/N_d$ lies in the range 2.37-6.25, the
probability that the model is consistent with the data is still between
10-50\%, so these classifications should be regarded as "tentative".

In Figure 2, we plot galaxies in the stellar mass range $10^{9.75}-10^{10}
M_{\odot}$ in the plane of log SFR/$M_*$ versus D$_n$(4000) and log
SFR/M$_*$ versus H$\delta_A$.  We colour-code the points according to their
classification: black refers to continuous, blue to ongoing burst and red
to past burst.  This particular mass range is chosen mainly for
illustrative purposes.  As can be seen, most of the leverage in the
classification comes from the combination of SFR/$M_*$ and D$_n$(4000)
where the three classes separate most clearly. This is not surprising,
because we have seen that these two quantities have smaller errors than
H$\delta_A$. It should also be noted that the class of galaxies that are
experiencing ongoing bursts is clearly separated from the the other two
classes in both planes. They appear as a very distinct cloud of points
extending to high values of SFR/$M_*$ and  low/high values of
D$_n$(4000)/H$\delta_A$.  The models in Figure 1 include many objects  with
underlying old stellar populations that are currently undergoing a burst.
These appear at high values of SFR/$M_*$ and high/low values of
D$_n$(4000)/H$\delta_A$. However, galaxies
of this type  do not appear to be present in the real data.

We comment briefly on the  very narrow tail of black points in the left panel of
Figure 2. This is the minority population of 
quiescent galaxies and/or AGN for which SFR is not estimated using emission
lines, but using D$_n$(4000). 
In the lower mass ranges, this tail disappears altogether.

\begin{figure}
\includegraphics[width=88mm]{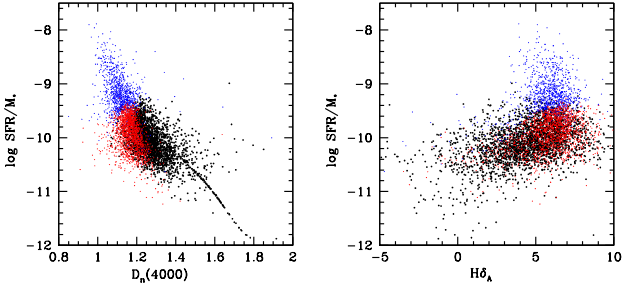}
\caption{Galaxies in the stellar mass range $10^{9.75}-10^{10}
M_{\odot}$ in the plane of log SFR/$M_*$ versus D$_n$(4000) and log
SFR/M$_*$ versus H$\delta_A$  We colour-code the points according to their
classification: black refers to continuous, blue to ongoing burst and red
to past burst. 
\label{examples}}
\end{figure}

We note that for each model galaxy, we store a variety of
parameters relating to its star formation history. We can then use the
library to compute the  most probable value of each parameter from the
median of the likelihood distribution, and an error on each parameter from
the range enclosing 68 percent of the total probability density (see the
Appendix in Kauffmann et al 2003a). The two parameters that will be explored
in this paper include the fraction of the total stellar mass of the galaxy
formed in the starburst and a formation time, defined to be the lookback
time when half the stars in the galaxy were formed.

\subsection {Evidence for IMF variations?} It is interesting to investigate
the population of galaxies with minimum reduced $\chi^2$ values greater
than 6.25, in order to see whether there is any evidence that  our
underlying stellar population models do not represent the data
adequately.  One possible
inadequacy is the assumption of a fixed Chabrier IMF. Changing the fraction
of high mass stars would alter the locus of the models in the plane of
inferred SFR/$M_*$ versus D$_n$(4000)/H$\delta_A$. Systematic variations in
the top-end of the IMF have been claimed by Hoversten \& Glazebrook (2008)
and Gunawardhana et al (2011) based on a changing relation between
H$\alpha$ equivalent width and broad-band colour, and by Meurer et al
(2009) based on a changing relation between H$\alpha$ and the UV
luminosities in galaxies.  We note that previous work did not attempt to
account for the effect of bursts on a galaxy-by-galaxy basis.  In addition,
the test using H$\alpha$-based SFR and two narrow-band stellar absorption
line indices should be less sensitive to systematic effects in the
procedures used to correct for dust extinction.

We find the fraction of galaxies with  $\chi^2_{min}/N_d$ values greater
than 6.25 to be $\sim$ 5\%. These outliers are primarily galaxies with low
4000 \AA\ break strengths and strong emission lines. The  greatest
discrepancy between models and data arises for the H$\delta_A$ index. We
note that in strongly star-bursting galaxies, the H$\delta_A$ absorption
feature is filled in by nebular emission. Although the MPA/JHU spectral
fitting pipeline attempts to subtract the emission, recovery of the
H$\delta$ absorption features can be expected to become  increasingly
inaccurate for strongly star-bursting galaxies where stellar features are
weak and nebular emission dominates. We therefore perform a restricted fit
to the combination of SFR/M$_*$ and D$_n$(4000). The percentage of galaxies
with $\chi^2_{min}/N_d$ values greater than 4.61 (corresponding to a
probability of 0.1 for 2 degrees of freedom) drops by another factor of two
to values less than $\sim$2-3\% in all the stellar mass bins.  The
remaining outliers mainly have D$_n$(4000) values that are less than 1, a
value that cannot be matched by the youngest stellar template in the
Bruzual \& Charlot (2003) models. We have checked the dependence of
D$_n$(4000) on IMF slope and on whether nebular continuum emission is
included using the STARBURST 99 models (Leitherer et al 1999).  Our tests
show that  D$_n$(4000)$< 1$ cannot be reached by tuning these parameters.
We think it is most likely that the remaining outliers are affected by
errors in the spectro-photometric flux calibration.  In summary, we find do 
not find a population of galaxies where there is conclusive evidence
for a top-heavy IMF, though we feel that it would be premature to rule
out IMF variations based on this data.

\section {Results} \subsection {Contribution of bursts to the SFR density}
The black curve in Figure 3 shows the fraction of the total integrated
stellar mass in galaxies that resides in galaxies that are well-fit by
continuous models.  The mass fraction increases from 0.3 for galaxies with
$M_*=10^8 M_{\odot}$ to 0.9 for galaxies with  $M_*= 10^{10.25} M_{\odot}$.
The blue curve shows the fraction of the total integrated star formation
rate density in galaxies with continuous star formation histories. This
rises from 0.05 at the low mass end to 0.7 at the high mass end. The cyan
curve shows the fraction of the total integrated star formation rate
density in galaxies currently undergoing a starburst. According to  Heckman et al (1997a),
at least 25\% of the high-mass star-formation in the local
universe occurs in starbursts, a number that agrees well with our results at
$M_*= 10^{10.25} M_{\odot}$. We note that the difference between
unity and the sum of the blue and cyan curves corresponds to the fraction
of the integrated SFR density in galaxies that are best fit by models with
past bursts. This generally lies in the range 0.1-0.2.

\begin{figure}
\includegraphics[width=78mm]{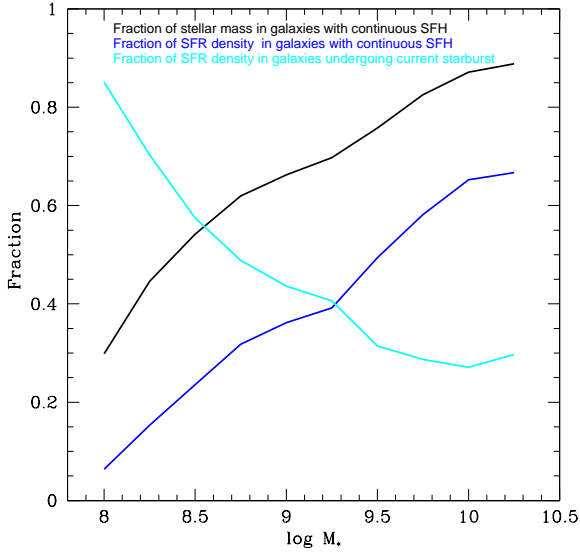}
\caption{The black curve  shows the fraction of the total integrated
stellar mass in galaxies that resides in galaxies that are well-fit by
continuous models. The blue curve shows the fraction of the total integrated star formation
rate density in galaxies with continuous star formation histories.
The cyan curve shows the fraction of the total integrated star formation rate
density in galaxies currently undergoing a starburst. 
\label{fcont}}
\end{figure}

\subsection {Burst mass fractions} In Figure 4, we plot distributions of
burst mass fractions for the population of galaxies currently undergoing
starbursts. The burst mass fraction is defined as the stellar mass
formed in the burst divided by the total stellar
mass of the galaxy.  Results are shown in 9 different mass ranges from 8.0 to 10.25
in log $M_*$. As can be seen, there is a significant tail of low mass
galaxies  with burst mass fractions larger than 0.3 for the lowest mass
bins. The tail of galaxies with large burst mass fractions disappears
progressively towards higher stellar masses. For galaxies with stellar
masses $\sim 10^{10} M_{\odot}$, the burst mass fraction is typically
around 0.05-0.1. The red dotted histograms in each panel shows the
distribution of the relative errors on $F_{burst}$, i.e. the error scaled by
dividing by the actual estimate. As can be seen, burst mass fractions can
be recovered with a typical accuracies of between 25 and 60\% (i.e.
significantly better than a factor of two). The error distribution does not
change significantly as a function of stellar mass, indicating that the
trend in the tail of strong starbursts is not an artifact of increasing
errors at low stellar masses.

\begin{figure}
\includegraphics[width=78mm]{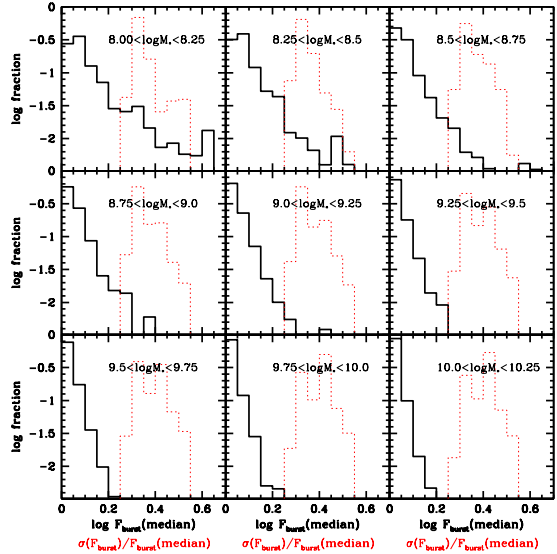}
\caption{Histograms (in black) of the distribution of the logarithm of the burst mass fraction.
Results are shown in 9 different mass ranges from 8.0 to 10.25
in log $M_*$. The red dotted histograms in each panel show the
distribution of the relative error on $F_{burst}$. 
\label{burststrength}}
\end{figure}

\subsection {Half mass formation times} In Figure 5, we plot distributions
of half-mass formation times for all galaxies in the same 9 stellar mass
ranges. Interestingly, in the lowest mass bin where the fraction of
galaxies experiencing ongoing starbursts is highest, the distribution of
half-mass formation times is very broad, ranging from 10 Gyr to less than 1
Gyr.  In the intermediate mass bins, the distribution is narrower, ranging
from 7 to 3 Gyr. In the highest mass bins, a tail of galaxies with large
($\sim$ 10 Gyr) formation times again appears. These results should be
compared to the distribution of 4000 \AA\ break strength shown in  Figure 2
of Kauffmann et al (2003b).  This plot shows a {\em monotonic} increase as a
function of stellar mass, with a secondary tail of large D$_n$(4000)
galaxies appearing at the same stellar mass where we see the large
half-mass formation time tail appear for the second time.

These results may appear puzzling at first sight, but we remind the reader
that a one-to-one mapping between 4000 \AA\ break strength and half mass
formation time only exists if the underlying star formation history is
continuous.  If star formation is bursty, D$_n$(4000) alone cannot be used
to infer formation time. The dotted red histograms in each panel show the
distribution of 1$\sigma$ errors on the formation time, which are in the
range 0.5-2.5 Gyr.  Once again the error distributions do not vary
significantly as a function of stellar mass, indicating that the greater
spread at low stellar masses is a real effect.

\begin{figure}
\includegraphics[width=78mm]{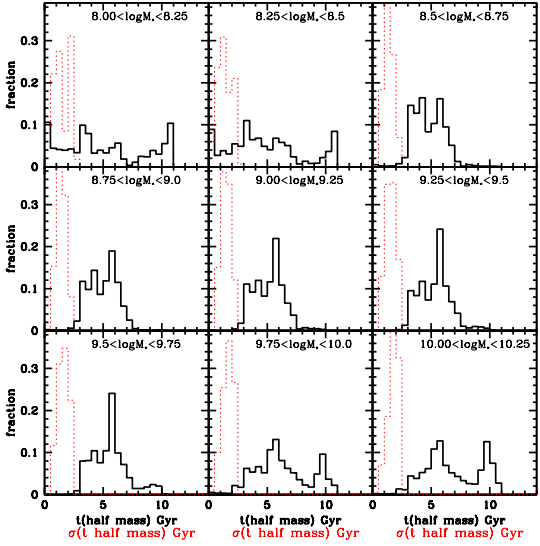}
\caption{Histograms (in black) of the distribution of the half mass formation time.                    
Results are shown in 9 different mass ranges from 8.0 to 10.25
in log $M_*$. The red dotted histograms in each panel show the
distribution of the  error on $t_{half mass}$. 
\label{formationtime}}
\end{figure}

\subsection {Correlation between burst mass fraction and galaxy structure
and metallicity} We now investigate whether there is any direct
observational evidence that bursts are correlated with changes in the
metallicities and structural properties of galaxies.  The structural
properties include the stellar surface mass density $\mu_*$, where $\mu_*=
0.5 M_*/ \pi R_{50}^2$ (R$_{50}$ is the half-light radius of the galaxy),
and the concentration index $R_{90}/R_{50}$, where $R_{90}$ is the radius
enclosing 90\% of the light of the galaxy. The concentration index is often
used as a rough proxy for galaxy bulge-to-disk ratio (Shimasaku et al
2001). The metallicity is the gas-phase metallicity derived from emission
line ratios using the methodology described in Tremonti et al (2004).

The solid curves in Figure 6 show how the median values of $\mu_*$, R90/R50
and gas-phase metallicity  vary as a function of $F_{burst}$ for galaxies
with ongoing starbursts. Dashed curves indicate the 25th and 75th
percentiles of the distribution of these quantities.  Results are shown in
three different mass ranges: $8< \log M_* < 9$, $9 < \log M_* < 10$ and $10 <
\log M_* < 10.5$. In the lowest mass bin, median stellar surface density
decreases by 0.2-0.3 dex in log $\mu_*$ for the highest burst mass
fractions.  No significant trend is seen for concentration index. The
gas-phase metallicity also drops by $\sim 0.2$ dex. In the two higher mass
bins, the correlation between stellar mass density and burst mass fraction
vanishes.  Instead, there is a clear trend for concentration index to
increase as a function of F$_{burst}$. The most natural interpretation of
these results is that in high mass galaxies, starbursts result in the
formation of a centrally-concentrated bulge component. In low mass
galaxies, starbursts no longer build bulges. The feedback associated with
the burst becomes strong enough to alter the stellar structure of the
galaxy and {\em reduce} stellar densities as seen in the hydrodynamical
simulations of dwarf galaxies discussed in Section 1.

\begin{figure}
\includegraphics[width=78mm]{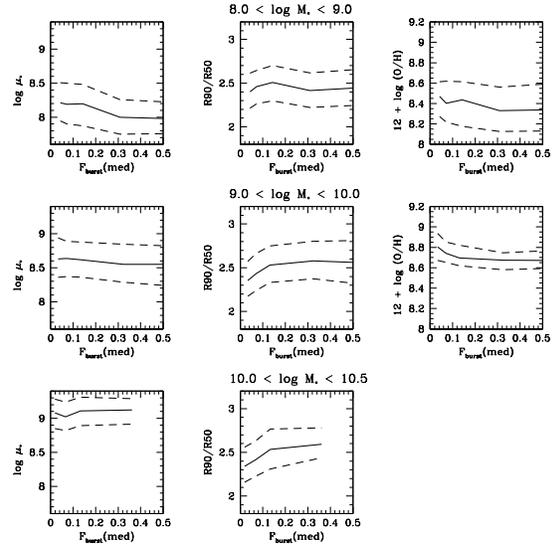}
\caption{Correlation between  $\mu_*$, R90/R50
and gas-phase metallicity as a function of $F_{burst}$.                    
Solid curves show the medians, while dashed curves indicate the 25th and 75th
percentiles of the distribution of each of these quantities.
\label{trends}}
\end{figure}

\subsection {Environment} We note that a reduction in the gas-phase
metallicities of interacting starburst galaxies in pairs compared to
``normal'' galaxies of the same stellar mass has been found in previous
work by Kewley, Geller \& Barton (2006). In Figure  7, we show an image
gallery of galaxies with stellar masses less than $10^9 M_{\odot}$ with
burst mass fractions larger than 0.4. As can be seen, the majority are
strongly asymmetric, but they are not always clearly interacting with
another nearby galaxy.  This is in agreement with work by Li et al. (2008),
who found that a close neighbour is a sufficient, but not necessary
condition for galaxies to be strongly star-forming.  We have also
investigated the environments of the star-bursting galaxies on larger
scales by counting the number of spectroscopic neighbours within a
projected radius of 1 Mpc and a velocity difference of less than 500 km/s.
We find no significant differences with respect to galaxies with continuous
star formation histories.

\begin{figure}
\includegraphics[width=85mm]{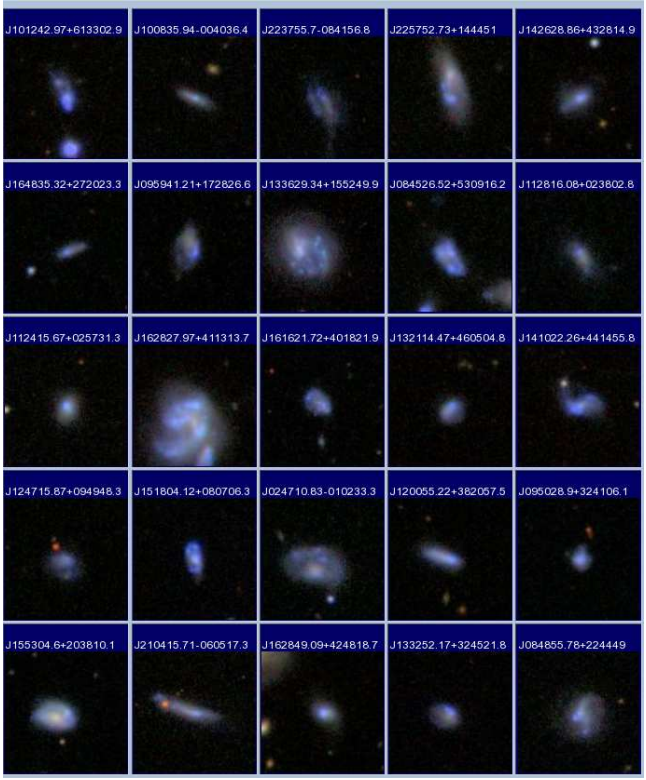}
\caption{SDSS  postage stamp images of some of the galaxies with burst mass fractions
greater than 0.4 in the stellar mass range $10^{8.5}-10^{9} M_{\odot}$.
\label{gallery_strongbursts}}
\end{figure}

\subsection {Burst amplitudes and durations} If we make the assumption that
starburst galaxies with similar stellar masses represent a single class of
object viewed a different times during the starburst cycle, the
distribution function of SFR/$M_*$ provides the most direct constraint on
the SFR amplitude variations during the burst.  In Figure 8, we plot the
cumulative distribution of galaxies with ongoing starbursts with  SFR/$M_*$
values less than a given value. Cyan, blue, green, black, yellow, red and
magenta curves represent galaxies in bins of increasing stellar mass from
$10^8$ to $10^{10.25} M_{\odot}$. Dotted lines mark the 10th and 90th
percentile points, which we adopt as our definition of peak-to-trough
variation in amplitude. As can be seen the peak-to-trough variations range
from a factor 25 for the lowest mass galaxies with $10^8 -10^9 M_{\odot}$
to a factor of 10 for galaxies with $\sim 10^{10} M_{\odot}$.

We can now combine these results from those presented in Figure 4 to
estimate the average duration of the bursts in low mass galaxies. The
median burst mass fraction in a galaxy with $10^8 M_{\odot}$ is 0.05, which
corresponds to a burst mass of $5 \times 10^6 M_{\odot}$. From Figure 8,
the median specific star formation rate during the burst is log SFR/$M_*$ =
-9.6, which corresponds to a star formation rate of 0.02511 $M_{\odot}$
yr$^{-1}$.  This leads to an estimate of the typical period of the burst of
$2 \times 10^8$ yr.  A $10^8 M_{\odot}$ galaxy has a characteristic
effective radius of 1.1 kpc (Kauffmann et al 2003b) and a rotation velocity
of 80 km s$^{-1}$ (McGaugh et al 2000), which yields a dynamical time of
$1.3 \times 10^8$ yr.  We conclude that the typical burst duration is very
similar to the dynamical timescale of the galaxy. This, together with our
estimates of peak-to-trough variation in SFR, are in good agreement with
the requirements put forward by Teyssier et al (2013).

\begin{figure}
\includegraphics[width=78mm]{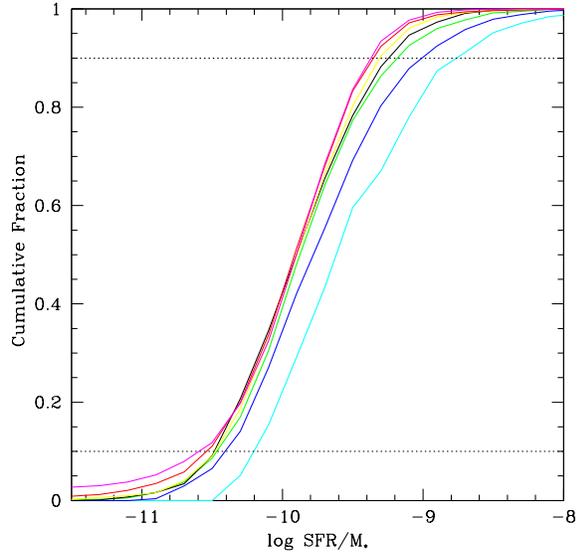}
\caption{ The
cumulative distribution of galaxies with ongoing starbursts with  SFR/$M_*$
values less than a given value. Cyan, blue, green, black, yellow, red and
magenta curves represent galaxies in bins of increasing stellar mass from
$10^8$ to $10^{10} M_{\odot}$. (Cyan: $10^8-10^{8.5} M_{\odot}$,
Blue: $10^{8.5}-10^{8.75} M_{\odot}$, Green:  $10^{8.75}-10^{9} M_{\odot}$,
Yellow:  $10^{9}-10^{9.25} M_{\odot}$, Red:  $10^{9.25}-10^{9.5} M_{\odot}$,
Magenta:  $10^{9.5}-10^{9.75} M_{\odot}$.)  Dotted lines mark the 10th and 90th
percentile points, which we adopt as our definition of "peak-to-trough"
variation in amplitude.
\label{cumul}}
\end{figure}

\section {Summary} We have used the the 4000 \AA\ break and H$\delta_A$
indices in combination with  SFR/$M_*$ derived from emission line flux
measurements, to constrain the recent star formation histories
of galaxies in the stellar mass range $10^{8}-10^{10} M_{\odot}$. Our main
results can be summarized as follows.

\begin {itemize}

\item The fraction of the total SFR density in galaxies with ongoing bursts
declines with increasing stellar mass, from 0.85 at a stellar mass of $10^8
M_{\odot}$ to 0.25 at a stellar mass of $10^{10} M_{\odot}$. The duty cycle
of bursts declines from 0.7 for  $10^8 M_{\odot}$ galaxies to 0.1 for
$10^{10} M_{\odot}$ galaxies.

\item The median burst mass fraction in galaxies of all stellar masses is
small, 5\% or less. There is, however, a tail of low mass galaxies that are
undergoing bursts that have formed as much as 50-60\% of their present-day
mass. The number  of galaxies in this high mass fraction tail decreases as
a function of stellar mass.

\item Low mass galaxies are not all young. The distribution of half mass
formation times for galaxies with stellar masses less than $10^9 M_{\odot}$ is
broad, spanning the range from 1 to 10 Gyr. This is quite different to
what is obtained when colours or 4000 \AA\ break strengths are used to
estimate stellar population age,  assuming continuous star formation
histories.

\item The peak-to-trough variation in star formation rate among the
bursting population ranges from a factor of 25 in the lowest mass galaxies
in our sample to a factor of 10 for galaxies with $M_*=10^{10} M_{\odot}$.
The average duration of bursts in low mass galaxies is comparable to their
average dynamical time.

\item Burst mass fraction is correlated with galaxy structure in quite
different ways in low and high mass galaxies. High
mass galaxies  experiencing strong bursts are more centrally
concentrated, indicating that bulge formation is likely at work. This is
not seen in low mass galaxies.  In low mass galaxies, we find that the
stellar surface densities decrease as a function of $F_{burst}$.

\item Gas phase metallicities decrease as a function of $F_{burst}$ in
galaxies of all stellar masses.

\end {itemize}

These results are in rather good agreement with the observational
predictions of Teyssier et al (2013) and lend further credence to the idea
that the cuspy halo problem can be solved by energy input from multiple
starbursts triggered by gas cooling and supernova feedback cycles over the
history of a low mass galaxy. We note that the analysis in this paper is
confined to the stellar populations enclosed within the 3 arsecond diameter
fiber aperture, which samples the central $\sim$ 20-30 \% of the
total light from the galaxy.  In future, it will be interesting to obtain
resolved  kinematic data for complete samples
of low mass galaxies to understand the influence of bursts on galaxy and
dark matter halo structure in more detail.

%===================================
\section*{Acknowledgments}
I thank Simon White for helpful discussions.

%===================================


\begin{thebibliography}{}

\bibitem[\protect\citeauthoryear{Abazajian et 
al.}{2009}]{2009ApJS..182..543A} Abazajian K.~N., et al., 2009, ApJS, 182, 
543 

\bibitem[\protect\citeauthoryear{Baldwin, Phillips, 
\& Terlevich}{1981}]{1981PASP...93....5B} Baldwin J.~A., Phillips M.~M., Terlevich R., 1981, PASP, 93, 5 

\bibitem[\protect\citeauthoryear{Bauer et al.}{2013}]{2013MNRAS.434..209B} 
Bauer A.~E., et al., 2013, MNRAS, 434, 209 

\bibitem[\protect\citeauthoryear{Brinchmann et 
al.}{2004}]{2004MNRAS.351.1151B} Brinchmann J., Charlot S., White S.~D.~M., 
Tremonti C., Kauffmann G., Heckman T., Brinkmann J., 2004, MNRAS, 351, 1151 

\bibitem[\protect\citeauthoryear{Bruzual 
\& Charlot}{2003}]{2003MNRAS.344.1000B} Bruzual G., Charlot S., 2003, MNRAS, 344, 1000 

\bibitem[\protect\citeauthoryear{Chabrier}{2003}]{2003PASP..115..763C} 
Chabrier G., 2003, PASP, 115, 763 

\bibitem[\protect\citeauthoryear{de Boer et 
al.}{2012}]{2012A&A...539A.103D} de Boer T.~J.~L., et al., 2012, A\&A, 539, A103 

\bibitem[\protect\citeauthoryear{Dolphin}{2012}]{2012ApJ...751...60D} 
Dolphin A.~E., 2012, ApJ, 751, 60 

\bibitem[\protect\citeauthoryear{Flores 
\& Primack}{1994}]{1994ApJ...427L...1F} Flores R.~A., Primack J.~R., 1994, ApJ, 427, L1 

\bibitem[\protect\citeauthoryear{Garrison-Kimmel et 
al.}{2013}]{2013MNRAS.433.3539G} Garrison-Kimmel S., Rocha M., 
Boylan-Kolchin M., Bullock J.~S., Lally J., 2013, MNRAS, 433, 3539 

\bibitem[\protect\citeauthoryear{Gnedin 
\& Zhao}{2002}]{2002MNRAS.333..299G} Gnedin O.~Y., Zhao H., 2002, MNRAS, 333, 299 

\bibitem[\protect\citeauthoryear{Governato et 
al.}{2010}]{2010Natur.463..203G} Governato F., et al., 2010, Natur, 463, 
203 

\bibitem[\protect\citeauthoryear{Gunawardhana et 
al.}{2011}]{2011MNRAS.415.1647G} Gunawardhana M.~L.~P., et al., 2011, 
MNRAS, 415, 1647 

\bibitem[\protect\citeauthoryear{Heavens et 
al.}{2004}]{2004Natur.428..625H} Heavens A., Panter B., Jimenez R., Dunlop 
J., 2004, Natur, 428, 625 

\bibitem[\protect\citeauthoryear{Heckman}{1997}]{1997AIPC..393..271H} 
Heckman T.~M., 1997, AIPC, 393, 271 

\bibitem[\protect\citeauthoryear{Hoversten 
\& Glazebrook}{2008}]{2008ApJ...675..163H} Hoversten E.~A., Glazebrook K., 2008, ApJ, 675, 163 

\bibitem[\protect\citeauthoryear{Kauffmann et 
al.}{2003}]{2003MNRAS.341...33K} Kauffmann G., et al., 2003a, MNRAS, 341, 33 

\bibitem[\protect\citeauthoryear{Kauffmann et 
al.}{2003}]{2003MNRAS.341...54K} Kauffmann G., et al., 2003b, MNRAS, 341, 54 

\bibitem[\protect\citeauthoryear{Kauffmann et 
al.}{2003}]{2003MNRAS.346.1055K} Kauffmann G., et al., 2003c, MNRAS, 346, 
1055 

\bibitem[\protect\citeauthoryear{Kewley, Geller, 
\& Barton}{2006}]{2006AJ....131.2004K} Kewley L.~J., Geller M.~J., Barton E.~J., 2006, AJ, 131, 2004 

\bibitem[\protect\citeauthoryear{Leitherer et 
al.}{1999}]{1999ApJS..123....3L} Leitherer C., et al., 1999, ApJS, 123, 3 

\bibitem[\protect\citeauthoryear{Li et al.}{2008}]{2008MNRAS.385.1903L} Li 
C., Kauffmann G., Heckman T.~M., Jing Y.~P., White S.~D.~M., 2008, MNRAS, 
385, 1903 

\bibitem[\protect\citeauthoryear{Li et al.}{2012}]{2012MNRAS.419.1557L} Li 
C., et al., 2012, MNRAS, 419, 1557 

\bibitem[\protect\citeauthoryear{Mashchenko, Wadsley, 
\& Couchman}{2008}]{2008Sci...319..174M} Mashchenko S., Wadsley J., Couchman H.~M.~P., 2008, Sci, 319, 174 

\bibitem[\protect\citeauthoryear{McGaugh et 
al.}{2000}]{2000ApJ...533L..99M} McGaugh S.~S., Schombert J.~M., Bothun 
G.~D., de Blok W.~J.~G., 2000, ApJ, 533, L99 

\bibitem[\protect\citeauthoryear{Meurer et al.}{1995}]{1995AJ....110.2665M} 
Meurer G.~R., Heckman T.~M., Leitherer C., Kinney A., Robert C., Garnett 
D.~R., 1995, AJ, 110, 2665 

\bibitem[\protect\citeauthoryear{Meurer et al.}{2009}]{2009ApJ...695..765M} 
Meurer G.~R., et al., 2009, ApJ, 695, 765 

\bibitem[\protect\citeauthoryear{Moore}{1994}]{1994Natur.370..629M} Moore 
B., 1994, Natur, 370, 629 

\bibitem[\protect\citeauthoryear{Navarro, Eke, 
\& Frenk}{1996}]{1996MNRAS.283L..72N} Navarro J.~F., Eke V.~R., Frenk C.~S., 1996, MNRAS, 283, L72 

\bibitem[\protect\citeauthoryear{Navarro, Frenk, 
\& White}{1996}]{1996ApJ...462..563N} Navarro J.~F., Frenk C.~S., White S.~D.~M., 1996, ApJ, 462, 563 

\bibitem[\protect\citeauthoryear{Pontzen 
\& Governato}{2012}]{2012MNRAS.421.3464P} Pontzen A., Governato F., 2012, MNRAS, 421, 3464 

\bibitem[\protect\citeauthoryear{Read 
\& Gilmore}{2005}]{2005MNRAS.356..107R} Read J.~I., Gilmore G., 2005, MNRAS, 356, 107 

\bibitem[\protect\citeauthoryear{Searle 
\& Sargent}{1972}]{1972ApJ...173...25S} Searle L., Sargent W.~L.~W., 1972, ApJ, 173, 25 

\bibitem[\protect\citeauthoryear{Shen et al.}{2013}]{2013ApJ...765...89S} 
Shen S., Madau P., Guedes J., Mayer L., Prochaska J.~X., Wadsley J., 2013, 
ApJ, 765, 89 

\bibitem[\protect\citeauthoryear{Shimasaku et 
al.}{2001}]{2001AJ....122.1238S} Shimasaku K., et al., 2001, AJ, 122, 1238 

\bibitem[\protect\citeauthoryear{Teyssier et 
al.}{2013}]{2013MNRAS.429.3068T} Teyssier R., Pontzen A., Dubois Y., Read 
J.~I., 2013, MNRAS, 429, 3068 

\bibitem[\protect\citeauthoryear{Tremonti et 
al.}{2004}]{2004ApJ...613..898T} Tremonti C.~A., et al., 2004, ApJ, 613, 
898 

\bibitem[\protect\citeauthoryear{Trujillo-Gomez et 
al.}{2013}]{2013arXiv1311.2910T} Trujillo-Gomez S., Klypin A., Colin P., 
Ceverino D., Arraki K., Primack J., 2013, arXiv, arXiv:1311.2910 

\bibitem[\protect\citeauthoryear{van der Wel et 
al.}{2011}]{2011ApJ...742..111V} van der Wel A., et al., 2011, ApJ, 742, 
111 

\bibitem[\protect\citeauthoryear{van Zee}{2001}]{2001AJ....121.2003V} van 
Zee L., 2001, AJ, 121, 2003 

\bibitem[\protect\citeauthoryear{York et al.}{2000}]{2000AJ....120.1579Y} 
York D.~G., et al., 2000, AJ, 120, 1579 

\bibitem[\protect\citeauthoryear{Zolotov et 
al.}{2012}]{2012ApJ...761...71Z} Zolotov A., et al., 2012, ApJ, 761, 71 


\end{thebibliography}
\end{document}